**Relation between grand canonical ensemble, Boltzmann, Fermi-Dirac, and Bose-Einstein distribution: Quantum principle for bosons and bosonic vacuum state, a candidate for dark energy and dark matter**

Markus Pollnau[*]
Advanced Technology Institute, University of Surrey, Guildford GU2 7XH, United Kingdom

[*]m.pollnau@surrey.ac.uk

**Abstract**
We obtain the two fundamental conditions that constitute a thermal equilibrium between two energy levels: (i) the total energy, i.e., the sum of kinetic and potential energy of all particles, must be equal in both levels and (ii) the temperature, hence the mean kinetic energy, must be equal for all particles. Exploiting these conditions, we derive a differential equation of thermal equilibrium that holds for all particles. Integration delivers the Boltzmann distribution, suggesting that it is the general distribution of thermal equilibrium. With excited-state and ground-state population numbers $n_2$ and $n_1$, respectively, Pauli's exclusion principle is formalized as $n_1 = 1 - n_2$. Exploiting Einstein's rate-equation approach to Planck's law of blackbody radiation, we derive the equivalent quantum principle for bosons, $n_1 = 1 + n_2$. Utilizing either quantum principle, as a boundary condition when integrating the differential equation of thermal equilibrium or by directly entering it into the Boltzmann distribution, delivers the Fermi-Dirac or Bose-Einstein distribution. The ratio of fermionic or bosonic population numbers, $n_2 / n_1$, follows the Boltzmann distribution. These results suggest that the Fermi-Dirac and Bose-Einstein distributions are special cases of the Boltzmann distribution, ruled by Pauli's exclusion principle or the quantum principle for bosons. The grand canonical ensemble is shown to be equivalent to the Boltzmann distribution. The general principle relating them, as well as the inhibition factor for fermions and the enhancement factor for bosons confirm the quantum principles. The quantum principle for bosons comprises a vacuum state. Consequently, a populated vacuum state should exist for all bosons. Vacuum states are not directly detectable; therefore, these states are dark states. The vacuum states of all bosons contribute to dark energy, while only the vacuum states of matter bosons contribute to dark matter, supporting the fact that the amount of dark energy exceeds that of dark matter in the universe.

**I. Introduction**

The Boltzmann distribution [1],

$$\frac{n_2}{n_1} = \frac{1}{e^{(E_2 - E_1)/k_B T}} , \qquad (1)$$

where $k_B$ is Boltzmann's constant and $T$ is the absolute temperature, hence $k_B T$ is the thermal energy, quantifies the population number $n_2$ of an energy level with energy $E_2$ with respect to the population number $n_1$ of a reference energy level with energy $E_1$ in thermal equilibrium at a given temperature $T$. In Section II, we will point out a few fundamental aspects of the Boltzmann distribution and the Boltzmann factors that are relevant to this work. The Boltzmann distribution can be derived in statistical physics by initially considering all possible population distributions and then obtaining the distribution of thermal equilibrium by maximizing the entropy [2–4].

Alternatively, the Boltzmann distribution has been derived—without considering statistical variations of the population numbers—by directly quantifying the situation of thermal equilibrium. In the chosen example, the atmospheric pressure decreases approximately exponentially with altitude [5]. The barometric formula [6] predicts an exponential decrease of atmospheric pressure with altitude, i.e., potential energy, and inverse temperature, thus representing a specific form of the Boltzmann distribution. We will briefly recall its derivation in Section III.

By generalizing this derivation in Section IV, we will obtain the two fundamental conditions that constitute a thermal equilibrium: (i) the total energy, i.e., the sum of potential and kinetic energy of all particles populating an energy level, must be equal in both energy levels, and (ii) the temperature, hence the mean kinetic energy of each individual particle, must be equal for all particles. Exploiting these two conditions yields the general differential equation of thermal equilibrium, whose integration delivers the Boltzmann distribution of Eq. (1). This route of deriving the Boltzmann distribution is not only significantly simpler than the statistical route but, moreover, provides insight that contrasts with conclusions that have been drawn from the grand canonical ensemble. Most importantly, since the two fundamental conditions of thermal equilibrium are independent of the nature of particles involved, the resulting Boltzmann distribution must apply to all particles, i.e., it must represent the general distribution of thermal equilibrium.

The natural consequence is that any other distribution of thermal equilibrium must necessarily be either equivalent to or a special case of the general Boltzmann distribution, obeying a specific condition. One can easily guess that Pauli's exclusion principle [7] is the specific condition fermions need to obey. In Section V, we will formalize Pauli's exclusion principle as $n_1 = 1 - n_2$, where $n_2$ represents the population number of the occupied or fermionic excited state, while $n_1$ represents the population number of the non-occupied or fermionic ground state. By exploiting Pauli's exclusion principle as a boundary condition when integrating the differential equation of thermal equilibrium, we obtain the Fermi-Dirac distribution [8–10],

$$n_{2,\text{FD}} = \frac{1}{e^{(E_2 - E_1)/k_B T} + 1} . \tag{2}$$

With the same quantum principle, we also derive the Fermi-Dirac distribution directly from the Boltzmann distribution, as well as the fermionic population numbers $n_1$ and $n_2$ from each other. Furthermore, the ratio $n_1 / n_2$ of fermionic excited- and ground-state population numbers follows the Boltzmann distribution and the fractional population numbers $n_1 / (n_1 + n_2)$ and $n_2 / (n_1 + n_2)$ equal the Boltzmann factors.

For reasons of symmetry, there must exist an equivalent quantum principle for bosons. In Section VI, we will utilize Einstein's semi-classical rate-equation approach [11] to Planck's law of blackbody radiation [12,13], which balances the spontaneous- and stimulated-emission rates with the absorption rate in an optical mode, in spectral resonance and thermodynamic equilibrium with the two-energy-level atomic oscillators supposedly existing in the walls of the blackbody radiator [11]. By introducing a bosonic excited state with population number $n_2$ and a bosonic ground state with population number $n_1$, we will derive the quantum principle for bosons, $n_1 = 1 + n_2$.

It will then be shown in Section VII that the quantum principle for bosons relates the Bose-Einstein distribution [14,15],

$$n_{2,\text{BE}} = \frac{1}{e^{(E_2 - E_1)/k_B T} - 1} , \tag{3}$$

and the Boltzmann distribution to each other in exactly the same manner as Pauli's exclusion principle relates the Fermi-Dirac and Boltzmann distributions to each other. By exploiting the quantum principle for bosons as a boundary condition when integrating the differential equation of thermal equilibrium, we obtain the Bose-Einstein distribution of Eq. (3). With the same quantum principle, we also derive the Bose-Einstein distribution directly from the Boltzmann distribution, as well as the bosonic population numbers $n_1$ and $n_2$ from each other. Furthermore, the ratio $n_1 / n_2$ of bosonic excited- and ground-state population numbers follows the Boltzmann distribution and the fractional population numbers $n_1 / (n_1 + n_2)$ and $n_2 / (n_1 + n_2)$ equal the Boltzmann factors. The relations between the Fermi-Dirac and Bose-Einstein distributions, on the one hand, and the Boltzmann distribution, on the other hand, lead us to the conclusion that fermions and bosons simultaneously obey both their own specific and the general Boltzmann distribution.

In Section VIII, we will point out that the relation between the Boltzmann distribution and the grand canonical ensemble,

$$n_{2,\text{GCE}} = \frac{1}{e^{(E_2 - E_1)/k_B T} + \eta} , \tag{4}$$

written here in the same notation as the other distributions, where $\eta$ is a principally free parameter, has been misinterpreted. We will show that the grand canonical ensemble and the Boltzmann distribution are equivalent, and both include the Fermi-Dirac and Bose-Einstein distributions as special cases. The general principle, $n_1 = 1 + \eta\, n_2$, that relates the grand canonical ensemble to the Boltzmann distribution straight-forwardly delivers and, thereby, confirms the quantum principles for fermions and bosons for $\eta = \pm 1$. By exploiting the general principle as a boundary condition when integrating the differential equation of thermal equilibrium, we obtain the grand canonical ensemble and the corresponding ground state. With the same general principle, we also derive the grand canonical ensemble directly from the Boltzmann distribution, as well as its population numbers $n_1$ and $n_2$ from each other. The ratio $n_1 / n_2$ of excited- and ground-state population numbers follows the Boltzmann distribution and the fractional population numbers $n_1 / (n_1 + n_2)$ and $n_2 / (n_1 + n_2)$ equal the Boltzmann factors.

The quantum-mechanical perspective taken in Section IX allows us to derive the quantum principles for fermions and bosons from the inhibition or enhancement factor of indistinguishable particles with antisymmetric (fermionic) or symmetric (bosonic) wavefunction, respectively.

Our increasingly detailed knowledge of the behavior of the smallest particles has had tremendous implications on our understanding of the universe. One of the great questions presently unsolved is about the nature of dark energy and dark matter [16–19]. In this respect, the quantum principle for bosons, $n_1 = 1 + n_2$, deserves our special attention. In the present paper, it is initially derived for photons, where the number one on the right-hand side represents the one vacuum photon per optical mode, i.e., the zero-point energy of the electromagnetic field in vacuum. However, this quantum principle is confirmed by the general principle that relates the grand canonical ensemble to the Boltzmann distribution, is also derived from the bosonic enhancement factor, and connects the Bose-Einstein and Boltzmann distributions to each other in a general manner; therefore, it must hold true for all bosons. Consequently, we will suggest in Section X that all bosonic distributions must comprise such a vacuum state of positive energy. Since this vacuum state is not directly detectable, it is a dark state. Foreseeably, the dark states of matterless bosons will contribute to dark energy, while the dark states of matter bosons will contribute simultaneously to dark matter and dark energy, thereby supporting the experimental evidence that the amount of dark energy exceeds that of dark matter in the universe [20].

## II. Boltzmann distribution and Boltzmann factors

At the beginning of this work poses a fundamental question. A beam balance with only one bar can never form an equilibrium. An equilibrium requires at least two physical quantities that are in equilibrium with each other. In the thermal equilibria described by Eqs. (1)–(4), the population numbers $n_1$ and $n_2$ of the two levels with energies $E_1$ and $E_2$ are in equilibrium with each other. The Boltzmann distribution of Eq. (1) includes these population numbers $n_1$ and $n_2$. In contrast, the Fermi-Dirac distribution, the Bose-Einstein distribution, and the grand canonical ensemble leave us in the dark about the population number $n_1$. It raises the question: what are the population numbers $n_1$ that are in thermal equilibrium with the population numbers $n_2$ of Eqs. (2)−(4)? We will answer this question during the present work. It will be the key to a deeper understanding of the thermal distributions and their relations.

Throughout this paper, when we discuss or calculate population numbers $n$ of energy levels, these are always understood as the mean population numbers, relating to the state of thermal equilibrium; for simplicity, we will leave out the angle brackets, $\langle n \rangle$, that are often used to indicate mean values. Furthermore, also for simplicity, we will assume non-degenerate energy levels $i$, i.e., their degeneracies are $g_i = 1$.

We will commence our investigation by briefly recalling a few fundamental aspects of the Boltzmann distribution of Eq. (1) that are relevant to this work. Usually, level 1 with a given energy $E_1$ is chosen as the reference level. The relative population number $n_2 / n_1$, which varies with energy $E_2$, leads to the graph of the Boltzmann distribution displayed in Fig. 1(a). When $E_2 = E_1$, then $n_2 / n_1 = n_1 / n_1 = 1$, i.e., the relative population number of level 1 is calibrated to unity. Equally well, one could choose level 2 with a fixed energy $E_2$ as the reference level and vary $E_1$, which only leads to a different calibration, as we will see in the example below.

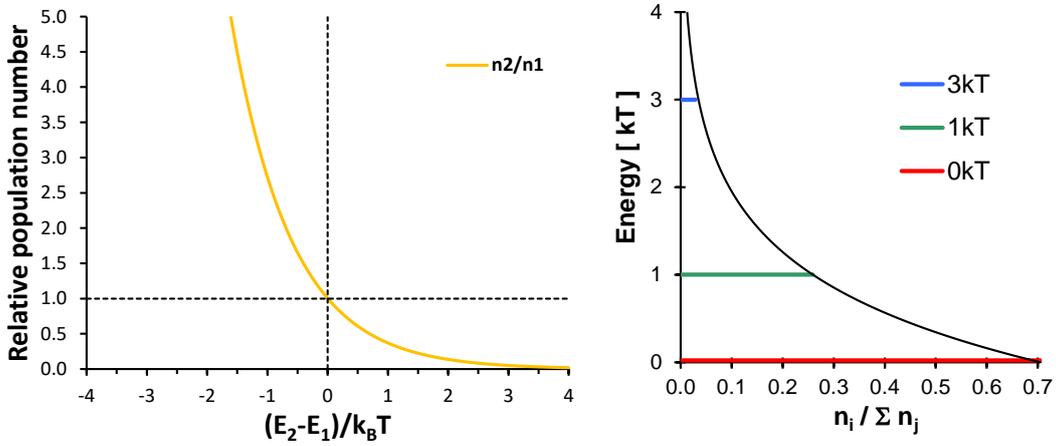

Fig. 1. (a) Boltzmann distribution of Eq. (1). Displayed is the relative population number $n_2 / n_1$ of a level 2 with varying energy $E_2$ versus a reference level 1 with fixed energy $E_1$. (b) Boltzmann factors of the example given in Table 1. The abscissa and ordinate in part (b) are reversed compared to part (a).

The Boltzmann factors $b_1$ and $b_2$ of two energy levels, i.e., their fractional population numbers, are

$$b_1 = \frac{n_1}{n_1 + n_2} = \frac{n_1}{n_1 e^{-(E_2-E_1)/k_B T} + n_1} = \frac{1}{e^{-(E_2-E_1)/k_B T} + 1} \tag{5}$$

and

$$b_2 = \frac{n_2}{n_1 + n_2} = \frac{n_1 e^{-(E_2-E_1)/k_B T}}{n_1 + n_1 e^{-(E_2-E_1)/k_B T}} = \frac{e^{-(E_2-E_1)/k_B T}}{1 + e^{-(E_2-E_1)/k_B T}} = \frac{1}{e^{(E_2-E_1)/k_B T} + 1}. \tag{6}$$

Extension to more than two energy levels leads to the Boltzmann factor of level $i$,

$$b_i = \frac{e^{-(E_i - E_1)/(k_B T)}}{\sum_j e^{-(E_j - E_1)/(k_B T)}}, \tag{7}$$

where the sum in the denominator is over all participating energy levels $j$. The Boltzmann factors are independent of the choice of reference level and their sum is normalized to unity, $\Sigma_i \, b_i = 1$.

An example is given in Table 1 and Fig. 1(b). We assume three levels with energies $E_1 = 0 \, k_B T$, $E_2 = 1 \, k_B T$, and $E_3 = 3 \, k_B T$. Either of the three levels is chosen as the reference level for the Boltzmann distribution, i.e.,

$$\begin{aligned}\frac{n_i}{n_1} &= \frac{1}{e^{(E_i - E_1)/k_B T}} \\ \frac{n_i}{n_2} &= \frac{1}{e^{(E_i - E_2)/k_B T}}, \\ \frac{n_i}{n_3} &= \frac{1}{e^{(E_i - E_3)/k_B T}}\end{aligned} \tag{8}$$

with $i = 1, 2, 3$. The Boltzmann distributions of Eq. (8) and the Boltzmann factors of Eq. (7) are calculated for each of the three choices of reference level. The relative population numbers calculated from the Boltzmann distribution and their sum change accordingly, because they are calibrated to different reference levels, whose Boltzmann distribution becomes unity. Independent of the choice of reference level, the fractional population numbers calculated as the Boltzmann factors remain the same and their sum is normalized to unity.

Table 1. Results of the Boltzmann distributions of Eq. (8) and the Boltzmann factors of Eq. (7) for three energy levels with energies $E_3 = 3 \, k_B T$, $E_2 = 1 \, k_B T$, and $E_1 = 0 \, k_B T$. In the three columns on the right-hand side, either of the three levels is chosen as the reference level for the Boltzmann distribution.

| Reference level | $n_{\text{ref}}$ | Level 1 | Level 2 | Level 3 |
|---|---|---|---|---|
| | $n_3 / n_{\text{ref}}$ | 0.050 | 0.135 | 1.000 |
| Boltzmann | $n_2 / n_{\text{ref}}$ | 0.368 | 1.000 | 7.389 |
| distribution | $n_1 / n_{\text{ref}}$ | 1.000 | 2.718 | 20.086 |
| | Sum | 1.418 | 3.854 | 28.475 |
| | $n_3 / \Sigma \, n_j$ | 0.035 | 0.035 | 0.035 |
| Boltzmann | $n_2 / \Sigma \, n_j$ | 0.259 | 0.259 | 0.259 |
| factors | $n_1 / \Sigma \, n_j$ | 0.705 | 0.705 | 0.705 |
| | Sum | 1.000 | 1.000 | 1.000 |

The Boltzmann distribution delivers relative population numbers $n_2 / n_1$, whereas the Boltzmann factors deliver fractional population numbers $n_i / (n_1 + n_2)$. Neither of them specifies the absolute population numbers $n_1$ and $n_2$ directly. Therefore, even when we have quantified the Boltzmann distribution and the Boltzmann factors, we do not know how many particles populate levels 1 and 2. A numerical example is given in Table 2. Assume that, for a given combination of $E_1$, $E_2$, and $T$, the value of the exponent in Eq. (1) equals $(E_2 - E_1) / k_B T = \ln(4)$, hence the Boltzmann distribution of Eq. (1) delivers $n_2 / n_1 = 0.25$, resulting in Boltzmann factors in Eqs. (5) and (6) of $b_1 = 0.8$ and $b_2 = 0.2$, respectively. Then infinitely many possible pairs of absolute population numbers $n_1$ and $n_2$, see the ten examples in columns 1–2 of Table 2, lead to the same ratio $n_2 / n_1$ and, therefore, fulfil these requirements, see columns 3–5 of Table 2. Needless to mention that one can create equivalent tables with infinitely many possible pairs of absolute population numbers $n_1$ and $n_2$ for any other ratio $n_2 / n_1$ resulting from the Boltzmann distribution. Columns 6–10 of Table 2 will be discussed in subsequent Sections.

To better understand the difference between absolute and relative population numbers, we can compare them to another physical situation. Assume that, instead of the absolute population numbers $n_1$ and $n_2$, we were dealing with the intensities $I_1$ and $I_2$ of light incident upon and transmitted through a light-absorbing dielectric medium, respectively. The Lambert-Beer law of absorption, $I_2 / I_1 = \exp(-\alpha \ell)$, where $\alpha$ is the absorption coefficient and $\ell$ is the length of the light-absorbing medium, delivers the relative transmitted light intensity $I_2 / I_1$. Whatever incident light intensity we choose, according to the Lambert-Beer law always the same relative light intensity $\exp(-\alpha \ell)$ is transmitted through the medium (in reality, this dependence holds only approximately, until the incident light intensity becomes so large that the ground state of the absorbing species starts to be bleached, an effect which is beyond the simplified physics considered by the Lambert-Beer law). I.e., there are infinitely many pairs of absolute light intensities $I_1$ and $I_2$ that result in the same relative light intensity $I_2 / I_1$. In this example, we understand already from their difference in physical units that one should not confuse the absolute light intensities $I_1$ and $I_2$ with the relative light intensity $I_2 / I_1$.

Table 2. Numerical example for the argument of exponent in the Boltzmann distribution of Eq. (1) equalling $(E_2 - E_1) / k_B T = \ln(4)$. For explanations, see the text.

| Absolute population numbers | | Boltzmann distribution, Eqs. (1), (5), (6) | | | Grand canonical ensemble, Eqs. (72), (76), (4) | | | General principle, Eq. (71) | Quantum distributions, Eqs. (2), (3) |
|---|---|---|---|---|---|---|---|---|---|
| $n_1$ | $n_2$ | $n_2 / n_1$ | $b_1$ | $b_2$ | $\eta$ | $n_{1,\text{CGE}}$ | $n_{2,\text{CGE}}$ | $n_1 = 1 - \eta\, n_2$ | |
| 0.5 | 0.125 | 0.25 | 0.8 | 0.2 | 4 | 0.5 | 0.125 | $n_1 = 1 - 4\, n_2$ | |
| 0.6667 | 0.1667 | 0.25 | 0.8 | 0.2 | 2 | 0.6667 | 0.1667 | $n_1 = 1 - 2\, n_2$ | |
| 0.8 | 0.2 | 0.25 | 0.8 | 0.2 | 1 | 0.8 | 0.2 | $n_1 = 1 - n_2$ | Fermi-Dirac |
| 0.8889 | 0.2222 | 0.25 | 0.8 | 0.2 | 0.5 | 0.8889 | 0.2222 | $n_1 = 1 - 0.5\, n_2$ | |
| 1 | 0.25 | 0.25 | 0.8 | 0.2 | 0 | 1 | 0.25 | $n_1 = 1$ | |
| 1.25 | 0.3125 | 0.25 | 0.8 | 0.2 | −0.8 | 1.25 | 0.3125 | $n_1 = 1 + 0.8\, n_2$ | |
| 1.3333 | 0.3333 | 0.25 | 0.8 | 0.2 | −1 | 1.3333 | 0.3333 | $n_1 = 1 + n_2$ | Bose-Einstein |
| 2 | 0.5 | 0.25 | 0.8 | 0.2 | −2 | 2 | 0.5 | $n_1 = 1 + 2\, n_2$ | |
| 4 | 1 | 0.25 | 0.8 | 0.2 | −3 | 4 | 1 | $n_1 = 1 + 3\, n_2$ | |
| 40 | 10 | 0.25 | 0.8 | 0.2 | −3.9 | 40 | 10 | $n_1 = 1 + 3.9\, n_2$ | |

### III. Barometric formula

The atmospheric pressure decreases approximately exponentially with increasing altitude [5]. Experimentally, the decrease in pressure is not exactly exponential, because the air temperature decreases slowly with increasing altitude and there occurs air flow in the atmosphere, resulting in local density fluctuations.

The simplest theoretical approach results in the isothermal barometric formula, which predicts an exponential decrease of air pressure with increasing altitude, described by the Boltzmann distribution [6]. In its derivation, one assumes, firstly, a constant temperature $T$ with varying altitude $h$, which is equivalent to the assumption of a constant temperature $T$ with varying energy $E$ in the derivation of the Boltzmann distribution. This condition is sometimes interpreted as the key requirement for a thermal equilibrium, but we will see below that it is only one of two mandatory conditions. One assumes, secondly, the absence of air flow, i.e., a situation of hydrostatic equilibrium, which means there are no density fluctuations around the equilibrium situation; it is equivalent to the assumption in the derivation of the Boltzmann distribution that a static state of maximum entropy that is quantified by the Boltzmann distribution can be reached. Therefore, the barometric formula is based on the same fundamental physical assumptions as the Boltzmann distribution, and both describe a situation of thermal equilibrium; it is not a coincidence in mathematical expression of two physically distinct phenomena.

In the derivation of the barometric formula, the sum of all forces, namely $F_{\text{bottom}}$ and $F_{\text{top}}$ due to gas pressure from the bottom and the top, respectively, and $F_G$ due to gravitation acting upon the mass in an infinitesimal air volume $A\,dh$ of area $A$ and height $dh$, must vanish:

$$F_{\text{bottom}} - F_{\text{top}} - F_G = 0 \quad \Rightarrow$$
$$pA - (p + dp)A - g\rho A\,dh = 0 \quad \Rightarrow . \tag{9}$$
$$dp = -\rho g\,dh$$

Here, $p$ or $p + dp$ is the atmospheric pressure at an altitude $h$ or $h + dh$, respectively, $g$ is the gravitational acceleration, and $\rho$ is the density of particles in the volume $A\,dh$. Had one considered that the density $\rho$ changes to $\rho + d\rho$, where $d\rho$ has a negative value, when moving from $h$ to $h + dh$, where $dh$ has a positive value, a second-order infinitesimal term $d\rho\,dh$ would have occurred in the derivation, which can be considered small enough to be negligible. By use of the ideal gas law [21],

$$p = \rho \frac{R_G}{M_{\text{mol}}} T , \tag{10}$$

where $R_G$ is the ideal gas constant and $M_{\text{mol}}$ is the molar mass, one can replace $\rho$ in Eq. (9) to obtain the differential equation

$$\frac{dp}{p} = -\frac{M_{mol} g}{R_G T} dh, \quad (11)$$

whose integration delivers the barometric formula,

$$\int_{p_1}^{p_2} \frac{dp}{p} = -\frac{M_{mol} g}{R_G T} \int_{h_1}^{h_2} dh \implies \frac{p_2}{p_1} = \frac{1}{e^{M_{mol} g (h_2 - h_1)/R_G T}}. \quad (12)$$

The barometric formula is a special example of the Boltzmann distribution of Eq. (1) for the case of pressure $p$ as a function of altitude $h$, as we will confirm below. Evidently, one can derive the equilibrium situation, the Boltzmann distribution, without considering statistical variations of the atmospheric pressure.

### IV. Differential equation of thermal equilibrium and Boltzmann distribution

The above approach can be generalized. In the following, we aim at replacing the pressure $p$ by the corresponding number $n$ of particles and the altitude $h$ by the corresponding potential energy $E$. By inserting the relations

$$\rho = \frac{m}{V}, \quad (13)$$

$$R_G = N_A k_B, \quad (14)$$

$$n = n_{mol} N_A = \frac{m N_A}{M_{mol}} \implies M_{mol} = \frac{m N_A}{n}, \quad (15)$$

where $m$ is the mass, $N_A$ is Avogadro's constant, and $n_{mol}$ is the number of mols, into the ideal gas law of Eq. (10), we obtain the kinetic energy $E_k$ of $n$ particles in an ideal gas at temperature $T$,

$$E_k = pV = n k_B T. \quad (16)$$

Furthermore, the potential energy of a mass $m$ at an altitude $h_i$ is

$$E_{p,i} = m g h_i. \quad (17)$$

Hence, the potential energy per particle, i.e., the energy $E_i$ of the energy level $i$ populated by the $n_i$ particles, is

$$E_i = \frac{E_{p,i}}{n_i} = \frac{m g h_i}{n_i}. \quad (18)$$

Inserting Eqs. (14)−(16) and (18) into Eq. (12) then confirms the physical equivalence of barometric formula and Boltzmann distribution,

$$\frac{p_2}{p_1} = \frac{p_2 V}{p_1 V} = \frac{n_2 k_B T}{n_1 k_B T} = \frac{n_2}{n_1}$$
$$\frac{1}{e^{M_{mol} g (h_2 - h_1)/R_G T}} = \frac{1}{e^{mg(h_2 - h_1) N_A / n N_A k_B T}} = \frac{1}{e^{(mgh_2/n - mgh_1/n)/k_B T}} = \frac{1}{e^{(E_2 - E_1)/k_B T}}. \quad (19)$$
$$\frac{p_2}{p_1} = \frac{1}{e^{M_{mol} g (h_2 - h_1)/R_G T}} \iff \frac{n_2}{n_1} = \frac{1}{e^{(E_2 - E_1)/k_B T}}$$

Since the barometric formula for the specific case of atmospheric pressure and the Boltzmann distribution are equivalent and the former can be derived directly, without considering statistical variations of the atmospheric pressure, it must be possible to derive also the Boltzmann distribution without considering statistical variations of the population numbers.

Firstly, since the temperature $T$ has been assumed constant, a change $dp$ in pressure $p$ and, therefore, a change $dE_k$ in kinetic energy $E_k$ in Eq. (16) can occur in a given volume $V$ due only to a change $dn$ in the number $n$ of particles,

$$dE_k = dp V = dn k_B T. \quad (20)$$

A change $dh$ in altitude $h$ results in a change in potential energy $E_p$ in Eq. (17) of

$$dE_p = mgdh. \tag{21}$$

Inserting Eq. (13), (20), and (21) into Eq. (9) then yields

$$\begin{aligned} \mathrm{d}pV &= -mg\mathrm{d}h \quad \Rightarrow \\ \mathrm{d}E_k &= -\mathrm{d}E_p \quad \Rightarrow \\ E_t &= E_k + E_p = \mathrm{const} \end{aligned} \tag{22}$$

The middle line of Eq. (22) demonstrates that, when stepping from the situation at altitude $h$ to that at altitude $h + \mathrm{d}h$, an increase in potential energy leads to an equivalent decrease in kinetic energy. Consequently, the sum of potential and kinetic energy, the total energy $E_t$, must be equal at both altitudes. Thereby, we have arrived at the two fundamental conditions that must hold to establish a thermal equilibrium between two energy levels. In both energy levels,

  (i)  the total energy $E_t$, i.e., the sum of kinetic and potential energy of all particles, must be equal, and
  (ii) the temperature $T$, hence the mean kinetic energy of each particle, must be equal.

Condition (i) ensures an energetic equilibrium, whereas adding condition (ii) results in a thermal equilibrium. Both together form the equivalent to the condition of maximum entropy.

Secondly, as we have already seen in Eq. (18), a particle in an energy level with energy $E$ has the potential energy $E$. For example, the discrete energy levels $E_i$ of electronic trajectories $i$ in Bohr's model of the hydrogen atom [22] are the potential energies due to the Coulomb attraction between electron and proton. Therefore, the potential energy $E_p$ of $n$ particles in this energy level is

$$E_p = nE. \tag{23}$$

A change $\mathrm{d}E$ in energy $E$ of this energy level results in all $n$ particles exhibiting this change, i.e., the potential energy $E_p$ of $n$ particles changes by

$$\mathrm{d}E_p = n\mathrm{d}E. \tag{24}$$

Inserting Eqs. (20) and (24) into the middle line of Eq. (22) yields

$$\mathrm{d}nk_BT = -n\mathrm{d}E \quad \Rightarrow \quad \frac{\mathrm{d}n}{n} = -\frac{\mathrm{d}E}{k_BT}. \tag{25}$$

This equation is the generalization of Eq. (11). The same result can be derived—more formally—directly from the fundamental conditions (i) and (ii) of a thermal equilibrium by inserting Eqs. (16), (20), (23), and (24),

$$\begin{aligned} E_t(E) &= E_t(E+\mathrm{d}E) \quad \Rightarrow \\ E_k(E) + E_p(E) &= E_k(E+\mathrm{d}E) + E_p(E+\mathrm{d}E) \quad \Rightarrow. \\ nk_BT + nE &= (n+\mathrm{d}n)k_BT + (n+\mathrm{d}n)(E+\mathrm{d}E) \end{aligned} \tag{26}$$

By choosing the origin of potential energy at $E = 0$,

$$nk_BT = (n+\mathrm{d}n)k_BT + (n+\mathrm{d}n)\mathrm{d}E, \tag{27}$$

resolving the brackets,

$$0 = \mathrm{d}nk_BT + n\mathrm{d}E + \mathrm{d}n\mathrm{d}E, \tag{28}$$

and neglecting the second-order infinitesimal term $\mathrm{d}n\mathrm{d}E$ (equivalent to the second-order infinitesimal term $\mathrm{d}\rho\mathrm{d}h$ mentioned in the previous Section), we obtain, again,

$$\frac{\mathrm{d}n}{n} = -\frac{\mathrm{d}E}{k_BT}. \tag{29}$$

This is the differential equation of thermal equilibrium. It relates the relative change $\mathrm{d}n/n$ in the number $n$ of particles to the change $\mathrm{d}E$ in potential energy $E$ of each particle in a thermal equilibrium, i.e., at constant temperature $T$ and constant mean kinetic energy $k_BT$ of each particle. From its derivation we understand the underlying physics. In a higher-energy level, each particle has a higher potential energy, whereas its kinetic energy remains constant, because the temperature remains constant; therefore, its total energy increases when elevated from a lower-energy to a higher-energy level. The only possible way of readjusting the total energy $E_t$ of Eq. (26) of all particles in the higher energy level is then by decreasing the number $n + dn$ of particles that occupy the higher-energy level. Therefore, d$n$ must be negative, if d$E$ is positive (and vice versa), as is the case in Eq. (29).

Since the only assumptions in its derivation are (i) the equivalence of total energy of both energy levels and (ii) the same temperature of all particles occupying these two energy levels to establish a thermal equilibrium, the differential equation of thermal equilibrium, Eq. (29), is universally valid and independent of the particles involved. Consequently, by simple integration we will derive from it the Boltzmann distribution, the Fermi-Dirac distribution, the Bose-Einstein distribution, and the grand canonical ensemble.

Integration of the differential equation of thermal equilibrium, Eq. (29), over the population numbers $n_1$ of an energy level with energy $E_1$ and $n_2$ of an energy level with energy $E_2$,

$$\int_{n_1}^{n_2} \frac{dn}{n} = -\frac{1}{k_B T} \int_{E_1}^{E_2} dE \, , \tag{30}$$

provides that the population numbers $n_2$ and $n_1$ of particles in the two levels in thermal equilibrium at temperature $T$ are related to each other by

$$\frac{n_2}{n_1} = \frac{1}{e^{(E_2 - E_1)/k_B T}} \, . \tag{31}$$

By simple integration of the differential equation of thermal equilibrium, we have obtained the Boltzmann distribution of Eq. (1). Evidently, by exploiting the conditions (i) and (ii) constituting a thermal equilibrium, it is possible to derive the equilibrium situation, the Boltzmann distribution, without considering statistical variations of the population numbers, also in its general form.

Since we have applied the most general boundary conditions to the integration in Eq. (30) and not imposed any specific restriction on the population numbers $n_1$ and $n_2$, Eq. (30) is the most general way of integrating the differential equation of thermal equilibrium, Eq. (29). Therefore, the Boltzmann distribution of Eq. (31) or (1) must necessarily be the most general distribution function that describes a thermal equilibrium. Any other distribution function describing a thermal equilibrium must be either equivalent to or a special case of the Boltzmann distribution of Eq. (31) or (1), and we will show in the following that this statement specifically applies to the Fermi-Dirac and Bose-Einstein distribution and the grand canonical ensemble.

### V. Pauli's exclusion principle and Fermi-Dirac Distribution

Fermions are half-integral-spin particles [23] that obey Pauli's exclusion principle [7]. Therefore, the maximum population number of fermions in a system is one, equivalent to the system being in its excited state with energy $E_2$. Alternatively, the system is not occupied by a fermion, i.e., it is empty, equivalent to the system being in its ground state with energy $E_1$. Situations of $E_2 > E_1$ or $E_2 < E_1$ are both possible, i.e., the expression "excited state" does not necessarily refer to a state of higher energy.

The mean population number $n_2$ at which the system is occupied by a fermion is given by the Fermi-Dirac distribution of Eq. (2). Consequently, the mean population number $n_1$ at which the system is empty must then be given by

$$n_{1,FD} = 1 - n_{2,FD} = 1 - \frac{1}{e^{(E_2 - E_1)/k_B T} + 1} = \frac{1}{e^{-(E_2 - E_1)/k_B T} + 1} \, . \tag{32}$$

Pauli's exclusion principle can, thus, be formalized as

$$n_{1,FD} = 1 - n_{2,FD} \, . \tag{33}$$

The system must be either occupied by a single fermion or empty, hence the population number $n_{i,FD}$ in an energy level $i$ can never exceed one.

We will now integrate the same differential equation of thermal equilibrium, Eq. (29). However, this time we additionally exploit Eq. (33), relating the population numbers $n_2$ and $n_1$ to each other by Pauli's exclusion principle, thereby creating a result that is more specific than the general Boltzmann distribution. The population number in the excited state is then given by

$$\int_{n_1=1-n_2}^{n_2} \frac{dn}{n} = -\frac{1}{k_B T} \int_{E_1}^{E_2} dE \quad \Rightarrow$$

$$\frac{n_2}{1-n_2} = e^{-(E_2-E_1)/k_B T} \quad \Rightarrow$$

$$n_2 = \frac{e^{-(E_2-E_1)/k_B T}}{1+e^{-(E_2-E_1)/k_B T}} \quad \Rightarrow \tag{34}$$

$$n_2 = \frac{1}{e^{(E_2-E_1)/k_B T}+1} = n_{2,\mathrm{FD}}$$

By simple integration of the differential equation of thermal equilibrium and exploiting Pauli's exclusion principle, we have obtained the Fermi-Dirac distribution of Eq. (2). By again exploiting Pauli's exclusion principle of Eq. (33), the population number in the ground state is obtained as

$$\int_{n_1}^{n_2=1-n_1} \frac{dn}{n} = -\frac{1}{k_B T} \int_{E_1}^{E_2} dE \quad \Rightarrow$$

$$\frac{1-n_1}{n_1} = e^{-(E_2-E_1)/k_B T} \quad \Rightarrow \tag{35}$$

$$n_1 = \frac{1}{e^{-(E_2-E_1)/k_B T}+1} = n_{1,\mathrm{FD}}$$

The sum $n_{1,\mathrm{FD}} + n_{2,\mathrm{FD}}$ equals unity, as required by Eq. (33). The two distributions and their sum are displayed in Fig. 2(a).

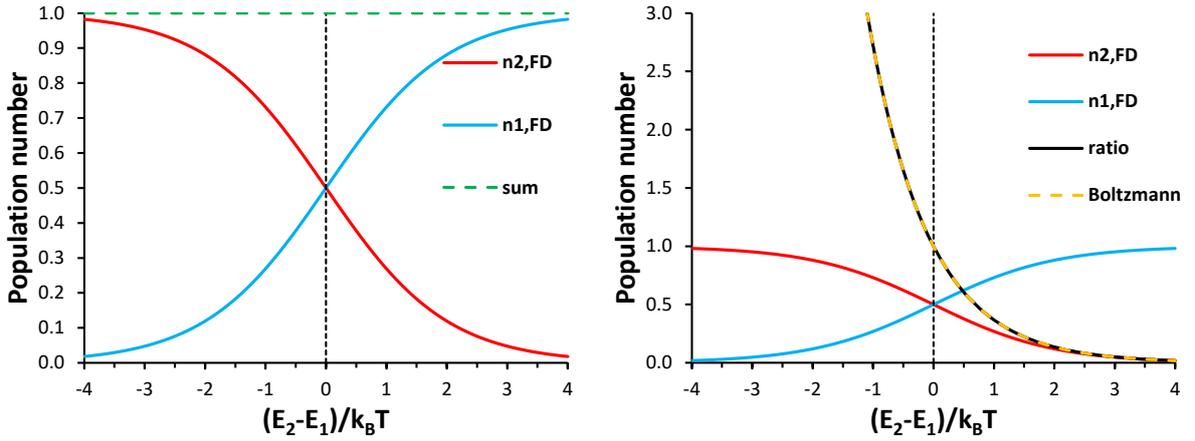

Fig. 2. Fermi-Dirac distribution $n_{2,\mathrm{FD}}$ of the occupied or fermionic excited state 2 of Eq. (34) and corresponding distribution $n_{1,\mathrm{FD}}$ of the empty or fermionic ground state 1 of Eq. (35). (a) Their sum, which equals unity. (b) Their ratio of Eq. (39), which equals the Boltzmann distribution of Eq. (31) or (1).

Both, the excited-state and ground-state population numbers of Eqs. (34) and (35), respectively, can also be derived by entering the quantum principle of Eq. (33) directly into the Boltzmann distribution of Eq. (31) or (1). For the excited state,

$$\frac{n_2}{n_1} = \frac{n_2}{1-n_2} = \frac{1}{e^{(E_2-E_1)/k_B T}} \quad \Rightarrow \tag{36}$$

$$n_2 = \frac{1}{e^{(E_2-E_1)/k_B T}+1} = n_{2,\mathrm{FD}}$$

I.e., entering Pauli's exclusion principle turns the Boltzmann distribution directly into the Fermi-Dirac distribution. The population number $n_1$ of Eq. (35) can be derived in the same manner by setting $n_2 = 1 - n_1$ in the Boltzmann distribution,

$$\frac{n_2}{n_1} = \frac{1-n_1}{n_1} = \frac{1}{e^{(E_2-E_1)/k_B T}} \quad \Rightarrow$$
$$n_1 = \frac{1}{e^{-(E_2-E_1)/k_B T}+1} = n_{1,\text{FD}} \quad . \tag{37}$$

Furthermore, Eqs. (34) and (35) can be derived from each other by use of the quantum principle of Eq. (33). For deriving $n_1$ from $n_2$, see Eq. (32); vice versa, $n_2$ can be derived from $n_1$ by

$$n_{2,\text{FD}} = 1 - n_{1,\text{FD}} = 1 - \frac{1}{e^{-(E_2-E_1)/k_B T}+1} = \frac{1}{e^{(E_2-E_1)/k_B T}+1} \quad . \tag{38}$$

The connection between the Boltzmann and the Fermi-Dirac distribution also holds in the opposite direction. Dividing the Fermi-Dirac distribution of Eq. (34) by the corresponding ground-state distribution of Eq. (35) results in the Boltzmann distribution of Eq. (31) or (1),

$$\frac{n_{2,\text{FD}}}{n_{1,\text{FD}}} = \frac{\dfrac{1}{e^{(E_2-E_1)/k_B T}+1}}{\dfrac{1}{e^{-(E_2-E_1)/k_B T}+1}} = \frac{1}{e^{(E_2-E_1)/k_B T}} = \frac{n_2}{n_1} \quad . \tag{39}$$

I.e., the ratio of the population numbers of fermionic excited and ground state equals the Boltzmann distribution, as displayed in Fig. 2(b).

Consequently, the same relation must necessarily hold also for the Boltzmann factors. The Fermi-Dirac distribution of Eq. (34) equals the Boltzmann factor $b_2$ of the excited state of Eq. (6),

$$\frac{n_{2,\text{FD}}}{n_{2,\text{FD}}+n_{1,\text{FD}}} = n_{2,\text{FD}} = \frac{1}{e^{(E_2-E_1)/k_B T}+1} = b_2 \quad . \tag{40}$$

Furthermore, the fermionic ground-state distribution of Eq. (35) equals the Boltzmann factor $b_1$ of the ground state of Eq. (5),

$$\frac{n_{1,\text{FD}}}{n_{2,\text{FD}}+n_{1,\text{FD}}} = n_{1,\text{FD}} = \frac{1}{e^{-(E_2-E_1)/k_B T}+1} = b_1 \quad . \tag{41}$$

The Fermi-Dirac distribution is the special case of the Boltzmann distribution for a single population, and the fractional Boltzmann factors $b_i$ normalize the Boltzmann distribution to exactly this single population, because Eqs. (5) and (6) divide the individual population number by the total population number $n_1 + n_2$.

We have found that Pauli's exclusion principle is the quantum principle for fermions that turns the general Boltzmann distribution into the more specific Fermi-Dirac distribution. Vice versa, the ratio of fermionic excited state, the Fermi-Dirac distribution, and fermionic ground state follows the Boltzmann distribution.

## VI. Quantum principle for bosons

For reasons of fundamental symmetry between half-integral-spin particles (fermions), which are quantum-mechanically described by an anti-symmetric wavefunction, and integral-spin particles (bosons) [23], which are quantum-mechanically described by a symmetric wavefunction, also the Bose-Einstein distribution must be a special case of the general Boltzmann distribution. Consequently, there must exist a quantum principle for bosons that is equivalent to Pauli's exclusion principle and turns the general Boltzmann distribution into the more specific Bose-Einstein distribution. Vice versa, there must exist a bosonic excited state, the Bose-Einstein distribution, and a corresponding bosonic ground state, whose ratio of population numbers follows the Boltzmann distribution.

Compared to Pauli's exclusion principle, deriving the quantum principle that bosons must fulfil is more involved [24], at least at our current level of understanding. We have exploited Einstein's rate-equation approach [11] to Planck's law of blackbody radiation [12,13] in a modern interpretation, i.e., by use of knowledge that was partly unavailable to Einstein in 1917. Einstein assumed that two-level atomic systems with level energies $E_2$ and $E_1$ are present in the walls of the blackbody radiator and that photons at a frequency $\nu$ are in resonance with the atomic transition,

$$h\nu = E_2 - E_1 \quad . \tag{42}$$

$h$ is Planck's constant, hence $h\nu$ is the photon energy. Photons are bosons, hence the mean number $\varphi$ of photons per optical mode follows the Bose-Einstein distribution of Eq. (3),

$$\varphi = \frac{1}{e^{h\nu/k_B T} - 1} . \tag{43}$$

The total photon energy per optical mode is

$$h\nu\varphi . \tag{44}$$

For a given spectral mode density $\tilde{M}(\nu)$ per unit frequency interval at frequency $\nu$, the spectral energy density $\tilde{u}(\nu)$ per unit frequency interval at frequency $\nu$ generated by the mean number $\varphi$ of photons per optical mode then amounts to

$$\tilde{u}(\nu) = \tilde{M}(\nu) h\nu\varphi . \tag{45}$$

When making the specific assumption that the optical system has the geometry of a simple three-dimensional box (or three-dimensional free space), which has a spectral mode density of $\tilde{M}(\nu) = 8\pi\nu^2/c^3$ [25], then inserting Eq. (43) into Eq. (45) delivers Planck's law of blackbody radiation [12,13],

$$\tilde{u}(\nu, T) = \frac{8\pi h\nu^3}{c^3} \frac{1}{e^{h\nu/k_B T} - 1} . \tag{46}$$

In the present derivation, we leave the mode structure and spectral mode density $\tilde{M}(\nu)$ of the optical system undefined, thereby allowing us to obtain a general result that holds true for any optical system.

The existence of a zero-point or vacuum energy, in optics manifested by vacuum photons, was proposed by Planck [26], further detailed by Einstein and Stern [27], and confirmed by Heisenberg [28]. Weisskopf suggested that spontaneous emission is triggered by this zero-point or vacuum energy [29–31]. Following this proposal, equivalently to Eq. (45), we interpret the vacuum spectral energy density $\tilde{u}_{vac}(\nu)$ in terms of a mean number $\varphi_{vac}$ of vacuum photons per optical mode,

$$\tilde{u}_{vac}(\nu) = \tilde{M}(\nu) h\nu\varphi_{vac} . \tag{47}$$

Consequently, the Einstein coefficients $A_{21}$ and $B_{21}$ of spontaneous and stimulated emission, respectively, are related by

$$A_{21} = B_{21} \tilde{u}_{vac}(\nu) . \tag{48}$$

Einstein's photon-rate equation [11] balances the spontaneous-emission rate $R_{sp}$ and the stimulated-emission rate $R_{st}$ with the absorption rate $R_{abs}$, in spectral resonance and thermodynamic equilibrium with a two-energy-level atomic system with population numbers $n_2$ and $n_1$ of upper and lower atomic level, respectively,

$$\begin{aligned} R_{sp} + R_{st} &= R_{abs} \quad \Leftrightarrow \\ A_{21} n_2 + B_{21} \tilde{u}(\nu) n_2 &= B_{12} \tilde{u}(\nu) n_1 \end{aligned}, \tag{49}$$

where $B_{12}$ is the Einstein coefficient of absorption. Inserting Eqs. (45), (47), and (48) into Eq. (49) yields

$$B_{21} \varphi_{vac} n_2 + B_{21} \varphi n_2 = B_{12} \varphi n_1 . \tag{50}$$

This is the photon-rate equation for a single optical mode, which is independent of the spectral mode density $\tilde{M}(\nu)$ of the physical system under consideration.

Einstein considered [11] that the atomic system is in thermal equilibrium, hence the atomic population numbers $n_2$ and $n_1$ are related by the Boltzmann distribution of Eq. (31) or (1). Inserting Eqs. (43) and (31) into Eq. (50) yields

$$\varphi_{vac}\left(1 - e^{-h\nu/k_B T}\right) = \frac{B_{12}}{B_{21}} - e^{-h\nu/k_B T} . \tag{51}$$

This equation holds true for all frequencies $\nu$ and all temperatures $T$ if

$$B_{12} = B_{21} . \tag{52}$$

As is well known, the Einstein coefficients of absorption and stimulated emission are equal [11] (under the condition that the degeneracies of the emitting and absorbing atomic energy levels are equal, as we have implicitly assumed in our derivation). Inserting Eq. (52) into Eq. (51) results in

$$\varphi_{\text{vac}} = 1. \tag{53}$$

It means that the Einstein coefficient or rate constant $A_{21}$ of spontaneous emission in Eq. (48) is induced by a vacuum spectral energy density that comprises one vacuum photon per optical mode in Eq. (47), as was already pointed out by Einstein and Stern [27] in 1913. Inserting the results of Eqs. (52) and (53) into Eq. (50) yields

$$n_2(\varphi + \varphi_{\text{vac}}) = n_2(\varphi + 1) = n_1 \varphi. \tag{54}$$

Equation (54) expresses the same essence as Eq. (50), namely that the rates of absorption, on the one hand, and spontaneous and stimulated emission, on the other hand, are equal in thermodynamic equilibrium. Although this derivation has possibly never been performed in the specific way presented here, its results are known until this point.

Consequently, the ratio of population numbers $n_2$ and $n_1$ of the two-level atomic system is in balance with the photon numbers $\varphi$ and $\varphi_{\text{vac}}$ in a single mode as

$$\frac{n_2}{n_1} = \frac{\varphi}{\varphi + \varphi_{\text{vac}}} = \frac{\varphi}{\varphi + 1}. \tag{55}$$

Since the left-hand side of this equation equals the Boltzmann distribution of Eq. (31) and its right-hand side comprises twice the Bose-Einstein distribution of Eq. (43), this equation relates the Boltzmann and Bose-Einstein distributions to each other. It is similar to Eq. (39), which relates the Boltzmann and Fermi-Dirac distributions to each other. Following the example of Eq. (39), we make the conceptual step of relating the Boltzmann population of Eq. (31) and the photon numbers in a single mode to the population numbers $n_{2,\text{BE}}$ and $n_{1,\text{BE}}$ of a photonic excited and ground state, respectively, which we introduce as

$$\frac{n_2}{n_1} = \frac{\varphi}{\varphi + \varphi_{\text{vac}}} = \frac{\varphi}{\varphi + 1} = \frac{n_{2,\text{BE}}}{n_{1,\text{BE}}}. \tag{56}$$

It leads to the relations

$$\begin{aligned} n_{2,\text{BE}} &= n_{\text{mode}} \varphi \\ n_{1,\text{BE}} &= n_{\text{mode}} (\varphi + \varphi_{\text{vac}}) = n_{\text{mode}} (\varphi + 1) \end{aligned}. \tag{57}$$

Mathematically, $n_{\text{mode}}$ is an arbitrary proportionality constant in the derivation of Eq. (57) from Eq. (56); physically, it is the number of modes, because the product $n_{\text{mode}} \varphi_{\text{vac}} = n_{\text{mode}}$ is the number of vacuum photons in a number $n_{\text{mode}}$ of modes. Setting $n_{\text{mode}} = 1$ yields

$$\begin{aligned} n_{2,\text{BE}} &= \varphi \\ n_{1,\text{BE}} &= \varphi + \varphi_{\text{vac}} = \varphi + 1 \end{aligned}. \tag{58}$$

In this manner, we have found the population numbers of the photonic excited and ground state of a single optical mode. The vacuum photon appears in the photonic ground state.

At first glance, the implication of Eq. (58) may look strange. How can the $\varphi$ photons in an optical mode simultaneously occupy an excited state with population number $n_{2,\text{BE}}$ and the ground state with population number $n_{1,\text{BE}}$? At this point we should recall that photons are virtual particles representing fields that convey interactions between real particles, and it is the interactions that matter here, not the virtual particles. Those fields or photons that trigger absorption belong to the excited state of the photonic system, whereas those fields or photons that trigger emission belong to the ground state of the photonic system. The vacuum field or photon triggers only emission, hence it appears only in the ground state, whereas real fields or photons trigger both absorption and emission, hence they appear in both states.

The difference in energy residing in the photonic excited and ground state is, thus, equal to the vacuum energy. Exploiting Eq. (58), we find that

$$n_{1,\text{BE}} = 1 + n_{2,\text{BE}}. \tag{59}$$

As we will see below, this is the quantum principle for photons and, more generally, all bosons. It is equivalent to Pauli's exclusion principle, the quantum principle for fermions of Eq. (33).

## VII. Bose-Einstein distribution

For the third time, we will now integrate the same differential equation of thermal equilibrium, Eq. (29). However, this time we additionally exploit Eq. (59), relating the absolute population numbers $n_2$ and $n_1$ to each other by the quantum principle for bosons, thereby making the result more specific than the general Boltzmann distribution and different from the Fermi-Dirac distribution. The population number in the excited state is then given by

$$\int_{n_1=1+n_2}^{n_2} \frac{dn}{n} = -\frac{1}{k_B T}\int_{E_1}^{E_2} dE \quad \Rightarrow$$
$$\frac{n_2}{1+n_2} = e^{-(E_2-E_1)/k_B T} \quad \Rightarrow$$
$$n_2 = \frac{e^{-(E_2-E_1)/k_B T}}{1-e^{-(E_2-E_1)/k_B T}} \quad \Rightarrow \qquad (60)$$
$$n_2 = \frac{1}{e^{(E_2-E_1)/k_B T}-1} = n_{2,BE}$$

By simple integration of the differential equation of thermal equilibrium and exploiting the quantum principle for bosons, we have obtained the Bose-Einstein distribution of Eq. (3). By again exploiting the quantum principle for bosons of Eq. (59), the population number in the ground state is obtained as

$$\int_{n_1}^{n_2=n_1-1} \frac{dn}{n} = -\frac{1}{k_B T}\int_{E_1}^{E_2} dE \quad \Rightarrow$$
$$\frac{n_1-1}{n_1} = e^{-(E_2-E_1)/k_B T} \quad \Rightarrow \qquad (61)$$
$$n_1 = \frac{1}{1-e^{-(E_2-E_1)/k_B T}} = n_{1,BE}$$

The difference $n_{1,BE} - n_{2,BE}$ equals unity, as required by Eq. (59). The two distributions and their difference are displayed in Fig. 3(a).

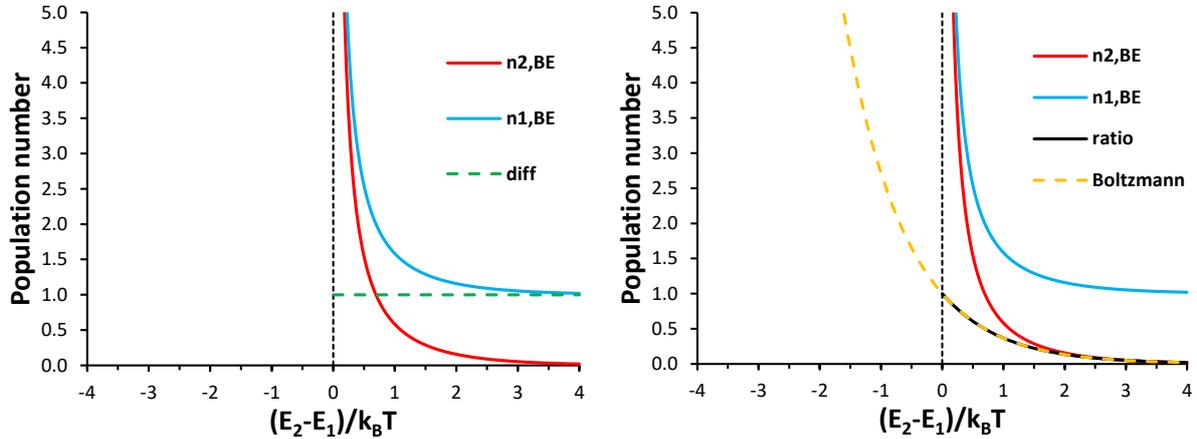

Fig. 3. Bose-Einstein distribution $n_{2,BE}$ of the bosonic excited state 2 of Eq. (60) and corresponding distribution $n_{1,BE}$ of the bosonic ground state 1 of Eq. (61). (a) Their difference, which equals unity. (b) Their ratio of Eq. (66), which equals the Boltzmann distribution of Eq. (31) or (1).

Both, the excited-state and ground-state population numbers of Eqs. (60) and (61), respectively, can also be derived by entering the quantum principle of Eq. (59) directly into the Boltzmann distribution of Eq. (31) or (1). For the excited state,

$$\frac{n_2}{n_1} = \frac{n_2}{n_2 + 1} = \frac{1}{e^{(E_2-E_1)/k_B T}} \quad \Rightarrow$$
$$n_2 = \frac{1}{e^{(E_2-E_1)/k_B T} - 1} = n_{2,BE} \tag{62}$$

I.e., entering the quantum principle for bosons turns the Boltzmann distribution directly into the Bose-Einstein distribution. The population number $n_1$ of Eq. (61) can be derived in the same manner by setting $n_2 = n_1 - 1$ in the Boltzmann distribution,

$$\frac{n_2}{n_1} = \frac{n_1 - 1}{n_1} = \frac{1}{e^{(E_2-E_1)/k_B T}} \quad \Rightarrow$$
$$n_1 = \frac{1}{1 - e^{-(E_2-E_1)/k_B T}} = n_{1,BE} \tag{63}$$

Furthermore, Eqs. (60) and (61) can be derived from each other by use of the quantum principle of Eq. (59):

$$n_{2,BE} = n_{1,BE} - 1 = \frac{1}{1 - e^{-(E_2-E_1)/k_B T}} - 1 = \frac{1}{e^{(E_2-E_1)/k_B T} - 1}, \tag{64}$$

$$n_{1,BE} = 1 + n_{2,BE} = 1 + \frac{1}{e^{(E_2-E_1)/k_B T} - 1} = \frac{1}{1 - e^{-(E_2-E_1)/k_B T}}. \tag{65}$$

The connection between the Boltzmann and the Bose-Einstein distribution also holds in the opposite direction. Dividing the Bose-Einstein distribution of Eq. (60) by the corresponding ground-state distribution of Eq. (61) results in the Boltzmann distribution of Eq. (31) or (1),

$$\frac{n_{2,BE}}{n_{1,BE}} = \frac{\frac{1}{e^{(E_2-E_1)/k_B T} - 1}}{\frac{1}{1 - e^{-(E_2-E_1)/k_B T}}} = \frac{1}{e^{(E_2-E_1)/k_B T}} = \frac{n_2}{n_1}. \tag{66}$$

I.e., the ratio of the population numbers of bosonic excited and ground state equals the Boltzmann distribution, as displayed in Fig. 3(b). This result has been expected, because we started from it in Eq. (56).

Then the calculation of fractional population numbers necessarily yields the Boltzmann factors of Eqs. (6) and (5),

$$\frac{n_{2,BE}}{n_{2,BE} + n_{1,BE}} = \frac{1}{\frac{n_{1,BE}}{n_{2,BE}} + 1} = \frac{1}{\frac{e^{(E_2-E_1)/k_B T} - 1}{1 - e^{-(E_2-E_1)/k_B T}} + 1} = \frac{1}{e^{(E_2-E_1)/k_B T} + 1} = b_2, \tag{67}$$

$$\frac{n_{1,BE}}{n_{2,BE} + n_{1,BE}} = \frac{1}{\frac{n_{2,BE}}{n_{1,BE}} + 1} = \frac{1}{\frac{1 - e^{-(E_2-E_1)/k_B T}}{e^{(E_2-E_1)/k_B T} - 1} + 1} = \frac{1}{e^{-(E_2-E_1)/k_B T} + 1} = b_1. \tag{68}$$

We have obtained the quantum principle for bosons, Eq. (59), and shown that it relates the Bose-Einstein distribution to the Boltzmann distribution in the exactly equivalent manner as Pauli's exclusion principle, Eq. (33), relates the Fermi-Dirac distribution to the Boltzmann distribution.

We can now obtain an overview of what the quantum principle of Eq. (33) for electrons and, more generally, fermions, i.e., Pauli's exclusion principle, and the quantum principle of Eq. (59) for photons and, more generally, bosons device. When integrating the differential equation of thermal equilibrium with general boundary conditions, we obtain the general Boltzmann distribution, whereas imposing either quantum principle on the boundary conditions during integration delivers the more specific Fermi-Dirac or Bose-Einstein distribution. When inserting either quantum principle into the Boltzmann distribution, we obtain the Fermi-Dirac or Bose-Einstein distribution, as well as the respective ground-state distribution. Exploiting either quantum principle in the Fermi-Dirac or Bose-Einstein distribution also yields the fermionic or bosonic ground-state distribution. The resulting relative population numbers $n_2 / n_1$ and fractional population numbers $n_i / (n_2 + n_1)$ of excited and ground state in the Fermi-Dirac and Bose-Einstein distributions obey the Boltzmann distribution and the Boltzmann factors, respectively. It means that, in a precisely equivalent manner, the two quantum principles transform the Boltzmann distribution into either quantum distribution, and vice versa. Apparently, the Fermi-Dirac and Bose-Einstein distributions can both be considered special cases of the Boltzmann distribution.

**VIII. Relation between grand canonical ensemble, Boltzmann, Fermi-Dirac, and Bose-Einstein distribution**

Of great practical importance is the grand canonical ensemble. It is often written in a notation for the mean population numbers $\langle n_i \rangle$ of levels $i$ with energies $E_i$ and sets the reference energy $E_1$ equal to the chemical potential $\mu$,

$$\langle n_i \rangle = \frac{1}{e^{(E_i - \mu)/k_B T} + \eta} . \tag{69}$$

Thereby, it relates the reference energy to the internal energy of a chemical substance of interest (e.g., a monoatomic ideal gas of a specific chemical element at normal atmospheric pressure) and, thus, interprets the thermal equilibrium distribution for countless chemical, physical, and engineering applications.

For direct comparison with the Boltzmann, Fermi-Dirac, and Bose-Einstein distribution of Eqs. (1)–(3), respectively, we will reset the chemical potential $\mu$ to the general reference energy $E_1$ and the mean population numbers $\langle n_i \rangle$ of levels $i$ with energies $E_i$ to the population number $n_2$ (in our chosen notation without angle brackets) of a single level with energy $E_2$, such that the grand canonical ensemble is represented by Eq. (4).

According to practically all the available literature on this subject, the specific cases of $\eta = 0$ and $\pm 1$ turn the grand canonical ensemble into the Boltzmann, Fermi-Dirac, and Bose-Einstein distribution, respectively. Since $\eta = 0$ and $\pm 1$ lead to three different equations, the Boltzmann, Fermi-Dirac, and Bose-Einstein distribution would be distinct from each other and, as is often claimed, the Fermi-Dirac and Bose-Einstein distribution could indeed be interpreted as quantum-mechanical corrections to the Boltzmann distribution for the cases of fermions and bosons, respectively. In contrast, in the previous Sections we have found that the Fermi-Dirac and Bose-Einstein distribution are special cases of the general Boltzmann distribution. Obviously, these interpretations disagree with each other.

Let us clarify this point. When inserting $\eta = \pm 1$ into Eq. (4), it becomes identical to Eq. (2) or (3), confirming that $\eta = \pm 1$ turns the grand canonical ensemble into the Fermi-Dirac or Bose-Einstein distribution. When inserting $\eta = 0$ into Eq. (4), the result compares with Eq. (1) in the following way:

$$\text{Grand canonical ensemble:} \quad n_2 = \frac{1}{e^{(E_2 - E_1)/k_B T}}$$
$$\text{Boltzmann distribution:} \quad \frac{n_2}{n_1} = \frac{1}{e^{(E_2 - E_1)/k_B T}} . \tag{70}$$

The right-hand sides are identical, but the left-hand sides are different. In comparative language studies, we would call the two objects on the right-hand side "false friends". Examples of false friends between English and French are the nouns "concurrence" (English) = coincidence versus "la concurrence" (French) = competition; or the adjectives "sensible" (English) = noticeable versus "sensible" (French) = sensitive. These words are the same in both languages but have different meanings. Similarly, the right-hand sides of Eq. (70) are the same, but they have different meanings, compare their left-hand sides. The grand canonical ensemble of Eq. (4), as well as the Fermi-Dirac and Bose-Einstein distributions of Eqs. (2) and (3), respectively, deliver the absolute population number $n_2$, whereas the Boltzmann distribution delivers the relative population number $n_2 / n_1$.

We have learned already in Section II that an absolute population number $n_2$ and a relative population number $n_2 / n_1$ are two very different physical quantities, and we have seen in Table 2 that, for a single relative population number $n_2 / n_1$, there exist infinitely many pairs of absolute population numbers $n_1$ and $n_2$ that fulfil this condition; obviously, $a = 0.25$ is not equal to $a / b = 0.25$. I.e., there should exist infinitely many results of the grand canonical ensemble of Eq. (4) that fulfil the Boltzmann distribution of Eq. (1), for the same value of the exponent [e.g., ln(4) in Table 2]. The choice of $\eta = 0$ in the grand canonical ensemble provides only one of these many solutions, namely the one for which $n_1 = 1$. There are infinitely many other values of $n_1$ and corresponding values of $n_2$ that fulfil the requirement of a single relative population number $n_2 / n_1$.

We find these infinitely many results of the grand canonical ensemble of Eq. (4) by equating it with Eq. (1). It provides the general principle that relates the grand canonical ensemble to the Boltzmann distribution,

$$n_1 = 1 - \eta n_2 . \tag{71}$$

If $\eta$ has been chosen and the population number $n_2$ has been calculated from the grand canonical ensemble, this general principle allows us to calculate the corresponding population number $n_1$ in the Boltzmann distribution. Vice versa, if the absolute population numbers $n_1$ and $n_2$ in the Boltzmann distribution are known, this general principle allows us to calculate the corresponding value of $\eta$ in the grand canonical ensemble,

$$\eta = \frac{1-n_1}{n_2}. \tag{72}$$

It means that the continuous range of $\eta$ values from infinity down to the negative value of $\eta$ that turns the denominator in Eq. (4) to zero is equal to the infinitely many possibilities of pairs of absolute population numbers $n_1$ and $n_2$ that fulfil the condition of a single relative population number $n_2 / n_1$. For a numerical example, see Table 2.

The general principle of Eq. (71) includes, as special cases, the quantum principles for fermions and bosons, respectively,

$$\eta = 1 \;\Leftrightarrow\; n_1 = 1 - n_2, \tag{73}$$

$$\eta = -1 \;\Leftrightarrow\; n_1 = 1 + n_2, \tag{74}$$

in agreement with the results of the previous Sections. It confirms that the Fermi-Dirac and Bose-Einstein distributions are special cases of the grand canonical ensemble (for $\eta = \pm 1$) and the Boltzmann distribution (for $n_1 = 1 \mp n_2$). For a numerical example, see Table 2 again.

For the fourth time now, we will exploit the fundamental conditions (i) and (ii) that constitute a thermal equilibrium by integrating the differential equation of thermal equilibrium of Eq. (29), this time with the boundary condition of the general principle of Eq. (71), to derive the grand canonical ensemble and the corresponding ground-state population number without considering statistical variations of the population numbers,

$$\begin{aligned}
\int_{n_1 = 1 - \eta n_2}^{n_2} \frac{dn}{n} &= -\frac{1}{k_B T} \int_{E_1}^{E_2} dE \;\Rightarrow\; \\
\frac{n_2}{1 - \eta n_2} &= e^{-(E_2 - E_1)/k_B T} \;\Rightarrow\; \\
n_2 &= \frac{e^{-(E_2 - E_1)/k_B T}}{1 + \eta e^{-(E_2 - E_1)/k_B T}} \;\Rightarrow\; \\
n_2 &= \frac{1}{e^{(E_2 - E_1)/k_B T} + \eta} = n_{2,\text{GCE}}
\end{aligned} \tag{75}$$

$$\begin{aligned}
\int_{n_1}^{n_2 = (1 - n_1)/\eta} \frac{dn}{n} &= -\frac{1}{k_B T} \int_{E_1}^{E_2} dE \;\Rightarrow\; \\
\frac{(1 - n_1)/\eta}{n_1} &= e^{-(E_2 - E_1)/k_B T} \;\Rightarrow\; \\
n_1 &= \frac{1}{\eta e^{-(E_2 - E_1)/k_B T} + 1} = n_{1,\text{GCE}}
\end{aligned} \tag{76}$$

See, again, Table 2 for a numerical example. Although we employ the general principle in this integration, the grand canonical ensemble is not a special case of the general Boltzmann distribution but is equivalent to it, because the general principle introduces the parameter $\eta$, which provides the necessary freedom. Setting in Eq. (76) $\eta = \pm 1$ provides the fermionic and bosonic ground states $n_{1,\text{FD}}$ and $n_{1,\text{BE}}$ of Eqs. (35) and (61), respectively. Finally, inserting the general principle into the Boltzmann distribution delivers the grand canonical ensemble, as was already clear from the way we derived the general principle of Eq. (71), one can derive the population numbers $n_{2,\text{GCE}}$ and $n_{1,\text{GCE}}$ from each other via the general principle,

$$n_{1,\text{GCE}} = 1 - \eta n_{2,\text{GCE}} = 1 - \eta \frac{1}{e^{(E_2 - E_1)/k_B T} + \eta} = \frac{1}{\eta e^{-(E_2 - E_1)/k_B T} + 1}, \tag{77}$$

$$n_{2,\text{GCE}} = \frac{1 - n_{1,\text{GCE}}}{\eta} = \frac{1 - \frac{1}{\eta e^{-(E_2 - E_1)/k_B T} + 1}}{\eta} = \frac{1}{e^{(E_2 - E_1)/k_B T} + \eta}, \tag{78}$$

their ratio,

$$\frac{n_{2,\text{GCE}}}{n_{1,\text{GCE}}} = \frac{\dfrac{1}{e^{(E_2-E_1)/k_B T}+\eta}}{\dfrac{1}{\eta e^{-(E_2-E_1)/k_B T}+1}} = \frac{1}{e^{(E_2-E_1)/k_B T}} = \frac{n_2}{n_1}, \tag{79}$$

equals the Boltzmann distribution, and the fractional population numbers of the grand canonical ensemble equal the Boltzmann factors of Eqs. (5) and (6),

$$\frac{n_{1,\text{GCE}}}{n_{1,\text{GCE}}+n_{2,\text{GCE}}} = \frac{\dfrac{1}{\eta e^{-(E_2-E_1)/k_B T}+1}}{\dfrac{1}{\eta e^{-(E_2-E_1)/k_B T}+1}+\dfrac{1}{e^{(E_2-E_1)/k_B T}+\eta}} = \frac{1}{e^{-(E_2-E_1)/k_B T}+1} = b_1, \tag{80}$$

$$\frac{n_{2,\text{GCE}}}{n_{1,\text{GCE}}+n_{2,\text{GCE}}} = \frac{\dfrac{1}{e^{(E_2-E_1)/k_B T}+\eta}}{\dfrac{1}{\eta e^{-(E_2-E_1)/k_B T}+1}+\dfrac{1}{e^{(E_2-E_1)/k_B T}+\eta}} = \frac{1}{e^{(E_2-E_1)/k_B T}+1} = b_2. \tag{81}$$

The obtained results demonstrate that the two conditions that constitute a thermal equilibrium, namely (i) the total energy, i.e., the sum of kinetic and potential energy of all particles, must be equal in both levels and (ii) the temperature must be equal for all particles, result in a differential equation of thermal equilibrium that describes a thermal equilibrium in its entirety. All relevant thermal distributions can be derived from it without first considering statistical variations of the population numbers and then deriving the equilibrium distribution by maximizing the entropy.

### IX. Indistinguishable particles, inhibition and enhancement factor, and quantum principles

The presence of particles in an energy level does not enhance or inhibit the probability of another distinguishable particle entering the same energy level. The situation changes when identical, indistinguishable particles are considered. $n$ indistinguishable particles in an energy level modify the probability of the $n + 1^{\text{st}}$ indistinguishable particle entering the same energy level by an inhibition factor of $1 - n$ for particles with an antisymmetric wavefunction (fermions), whose amplitudes add with a minus sign, or an enhancement factor of $1 + n$ for particles with a symmetric wavefunction (bosons), whose amplitudes add with a plus sign [32].

The principle of detailed balancing states that, in equilibrium, each elementary process is in equilibrium with its reverse process [33,34]. Einstein exploited this principle in his semi-classical rate-equation approach to Planck's law of blackbody radiation to derive the photon (Bose-Einstein) distribution [11], and Ornstein and Kramers utilized it in their discussion of the Fermi-Dirac distribution [35].

In the inhibition factor of $1 - n$ for fermions, we have the freedom to interpret $n$ as $n_1$ or $n_2$, and the same intuitive argument as in Section V leads to

$$n_1 = 1 - n_2, \tag{82}$$

which is Pauli's exclusion principle, the quantum principle for fermions of Eq. (33). The population numbers of the two energy levels must be connected by Eq. (82) to fulfil the requirement that, in equilibrium, the transition rates in both directions are equal.

For the example of photons, one can understand from the classical properties of interference and the requirement of energy conservation that the inverse transitions are the emission (addition) of one photon to a mode populated by $n$ photons and the absorption (removal) of one photon from a mode populated by $n + 1$ photons [36,37]. More generally, the same holds for all bosons [32]. The transition amplitudes are given by [32]

$$\begin{aligned}\langle n+1 | n \rangle &= \sqrt{n+1}\, a \\ \langle n | n+1 \rangle &= \sqrt{n+1}\, a^*\end{aligned}, \tag{83}$$

where $a$ and its complex conjugate $a^*$ are the transition amplitudes when no other bosons are present. Consequently, the transition probabilities of adding one boson to an energy level populated by $n$ bosons and the inverse probability of removing one boson from an energy level populated by $n + 1$ bosons are equal,

$$P = \left|\sqrt{n+1}\, a\right|^2 = \left|\sqrt{n+1}\, a^*\right|^2 = (n+1)|a|^2, \tag{84}$$

and enhanced by the same factor of $n + 1$ compared to the probabilities, $P = |a|^2$, when the particles are distinguishable. In the Bose-Einstein distribution we always find $E_2 > E_1$, see Fig. 3, and in thermal equilibrium the level with lower energy is always more populated; hence, we always find $n_1 > n_2$. To ensure that the transition rates are the same in both directions, the two population numbers must then be related by

$$n_1 = 1 + n_2, \tag{85}$$

which is the quantum principle for bosons of Eq. (59). Quantum-mechanically, the inhibition factor for indistinguishable fermions and the enhancement factor for indistinguishable bosons lead to the same quantum principles that we have derived in the previous Sections.

Oster derived from the principle of detailed balancing, albeit by introducing the thermal energy $k_B T$ in a non-intuitive manner, the Boltzmann distribution [38]. He overlooked the fact that it is the general distribution of thermal equilibrium, although he derived the Fermi-Dirac and Bose-Einstein distribution by adding the inhibition and enhancement factor [38].

Kaniadakis and Quarati proposed that the parameter $\eta$ in the grand canonical ensemble reflects the degree of indistinguishability of the particles, ranging from complete distinguishability for $\eta = 0$ to complete indistinguishability for $\eta = \pm1$ [39]. The question which situations the parameter values of $\eta > 1$ (beyond Fermi-Dirac) and $\eta < -1$ (beyond Bose-Einstein) in Table 2 describe remains open. These are neither fermionic nor bosonic states; possibly, these situations do not occur in nature, even if the Boltzmann distribution and the grand canonical ensemble do not exclude them.

The quantum-mechanical perspective concludes our comparison between the grand canonical ensemble, the Boltzmann, Fermi-Dirac, and Bose-Einstein distribution. In this mosaic, all pieces fit together and form a complete picture. Had we started our investigation by either comparing the grand canonical ensemble with the Boltzmann distribution or exploiting the quantum-mechanical situation of indistinguishable particles, we would have obtained the quantum principles for fermions and bosons from the general principle of Eq. (71) by Eqs. (73) and (74) or from the fermionic inhibition and bosonic enhancement factors by Eqs. (82) and (85). Hence, we could have avoided the effort made in Section VI to find the quantum principle for bosons via Einstein's rate-equation approach to Planck's law of blackbody radiation. Nevertheless, understanding the physics behind the quantum principle for bosons in Section VI will assist us now in making an interesting discovery.

## X. Bosonic vacuum state: A candidate for dark energy and dark matter

The existence of dark energy and dark matter in the universe has been evident for many decades [16–19]. Nonetheless, the physical origin of dark energy and dark matter has been an enigma ever since their discovery. Current estimations predict that the total mass-energy content includes ~5% of ordinary matter, ~27% of dark matter, and ~68% of other dark energy [20], i.e., ~95% of the total energy (including mass-energy) has as yet not been identified.

Let us return to the quantum principle for bosons of Eq. (59). It is clear from its derivation from Eq. (58) that the number 1 on its right-hand side is, in the case of electromagnetic energy, equivalent to the one vacuum photon per optical mode,

$$n_{1,BE} = \varphi_{vac} + n_{2,BE}. \tag{86}$$

Detection of this vacuum energy would require its conversion to, e.g., an electric signal that can be read out from a detector and then counted or displayed. As a result of this energy-conversion process, the photonic energy of the optical mode would need to jump to an even lower energy level. Unfortunately, an energy level lower than the zero-point energy of vacuum does not exist, hence the photonic vacuum energy cannot be converted and, therefore, cannot be directly detected. However, it reveals its existence, e.g., by the spontaneous emission it triggers without being consumed itself.

Since the Bose-Einstein distribution describes the behavior of all bosons and the quantum principle of Eq. (59) relates the Boltzmann and Bose-Einstein distributions to each other in a general manner, this quantum principle must hold true for all bosons, which has been confirmed by the comparison with the grand canonical ensemble and the quantum-mechanical perspective. It implies that not only photons but all bosons comprise an occupied vacuum state with a positive zero-point energy and population number

$$n_{vac,BE} = 1, \tag{87}$$

such that, in an extension of Eq. (86) to all bosons, the quantum principle for bosons of Eq. (59) reads

$$n_{1,BE} = n_{vac,BE} + n_{2,BE}. \tag{88}$$

There is a simple argument that supports this view. The number 1 in the enhancement factor of bosons represents the classical transition amplitudes $a$ and $a^*$ in Eq. (83). If additional transition amplitudes are mediated by additional bosons, also the classical transition rate should be mediated by a boson, namely the vacuum boson, like the vacuum photon triggers spontaneous emission. Then each boson, including the vacuum boson, induces the same probability of another boson entering the same energy level, no matter how many bosons are present, which we know is exactly the case for photons and the vacuum photon, see Eqs. (45), (47), (48), and (50).

In all situations, it will be impossible to directly detect this zero-point energy or vacuum boson. Nevertheless, it will trigger action in the universe, as does the vacuum photon. Since this zero-point energy is not directly detectable, it is a dark state.

The bosonic vacuum state of matterless bosons will only contribute to dark energy. In contrast, many bosons have a mass. Foreseeably, the mass-energy of their vacuum state will not be directly detectable, either. Hence, the bosonic vacuum state of matter bosons will simultaneously contribute to dark matter and dark energy. Consequently, the existence of bosonic vacuum states of matterless and matter bosons will naturally contribute to the difference between dark energy (including mass-energy) and dark matter in the universe.

## XI. Conclusions

We have learned from the derivation and nature of the barometric formula that it is not necessary to consider statistical variations of a thermal distribution to find the equilibrium distribution. Generalizing its derivation has led us to the two conditions that constitute a thermal equilibrium. Combining these two conditions has provided the differential equation of thermal equilibrium, $dn/n = -dE/k_B T$, whose integration has delivered the Boltzmann distribution. Since no assumptions about participating particles were made until this point, the Boltzmann distribution holds for all particles. Imposing a specific condition, namely either Pauli's exclusion principle for fermions, $n_1 = 1 - n_2$, or the corresponding quantum principle for bosons, $n_1 = 1 + n_2$, which we have derived for photons from Einstein's rate-equation approach to Planck's law of blackbody radiation, has delivered the Fermi-Dirac and Bose-Einstein distribution, respectively. All relations between these three distributions obtained in the present work identify the Fermi-Dirac and Bose-Einstein distribution as special cases of the general Boltzmann distribution. Revisiting the grand canonical ensemble has shown that it is equivalent to the Boltzmann distribution and has confirmed these results. Furthermore, we have confirmed the two quantum principles via the relation between the grand canonical ensemble and the Boltzmann distribution, as well as from the quantum-mechanical perspective.

The quantum principle obtained for bosons, $n_1 = 1 + n_2$, has significant implications. From its derivation and its validity for all bosons, we can predict that not only the photonic distribution, but all bosonic distributions must comprise a populated vacuum state of positive energy. Since a vacuum state is not directly detectable and, therefore, represents a dark state, all bosonic distributions will contribute to dark energy, and distributions of matter bosons will also contribute to dark matter. Whether this finding solves the enigma surrounding the physical nature of dark energy and dark matter in the universe in its entirety or possibly makes only a small contribution to the explanation awaits further investigations.

## Acknowledgments

The author gratefully acknowledges discussions with (in chronological order) Marc Eichhorn, Jeremy Allam, Nur Ismail, Konstantin Litvinenko, and Rafael Valiente. This work was partially funded by the European Research Council via the ERC Advanced Grant No. 341206.